\documentclass[reprint,showpacs,twocolumn,superscriptaddress,aps,prx]{revtex4-2}

\usepackage[utf8]{inputenc}
\usepackage{bbm}
\usepackage{bm}
\usepackage[usenames]{color}
\usepackage{multirow}
\usepackage{amssymb}
\usepackage{amsbsy}
\usepackage{mathtools}
\usepackage{amsmath}
\usepackage{stmaryrd}
\usepackage{graphicx}
\usepackage{soul}
\usepackage{epsfig}
\usepackage{placeins}
\usepackage{bbold}
\usepackage{braket}
\usepackage{blindtext}
\usepackage[colorlinks,linkcolor=blue,citecolor=blue,urlcolor=blue]{hyperref}

\usepackage{scalerel}
\usepackage{tikz}
\usetikzlibrary{calc}
\usetikzlibrary{patterns}
\usetikzlibrary{svg.path}
\definecolor{orcidlogocol}{HTML}{A6CE39}
\tikzset{
  orcidlogo/.pic={
    \fill[orcidlogocol] svg{M256,128c0,70.7-57.3,128-128,128C57.3,256,0,198.7,0,128C0,57.3,57.3,0,128,0C198.7,0,256,57.3,256,128z};
    \fill[white] svg{M86.3,186.2H70.9V79.1h15.4v48.4V186.2z}
                 svg{M108.9,79.1h41.6c39.6,0,57,28.3,57,53.6c0,27.5-21.5,53.6-56.8,53.6h-41.8V79.1z M124.3,172.4h24.5c34.9,0,42.9-26.5,42.9-39.7c0-21.5-13.7-39.7-43.7-39.7h-23.7V172.4z}
                 svg{M88.7,56.8c0,5.5-4.5,10.1-10.1,10.1c-5.6,0-10.1-4.6-10.1-10.1c0-5.6,4.5-10.1,10.1-10.1C84.2,46.7,88.7,51.3,88.7,56.8z};
  }
}

\newcommand\orcid[1]{\!%
  \href{https://orcid.org/#1}{%
    \mbox{%
      \scaleto{%
        \begin{tikzpicture}[yscale=-1,transform shape]
          \pic{orcidlogo};
        \end{tikzpicture}
      }{8pt}%
    }%
  }%
}

\begin{document}

\title{Lanczos-Pascal approach to correlation functions in chaotic quantum systems
}

\author{Merlin F\"{u}llgraf~\orcid{0009-0000-0409-5172}
}
\email{merlin.fuellgraf@uos.de}
 \author{Jiaozi Wang
\orcid{0000-0001-6308-1950}
 }
 \author{Robin Steinigeweg
\orcid{0000-0003-0608-0884}
 }

\author{Jochen Gemmer
\orcid{0000-0002-4264-8548}
}
\email{jgemmer@uos.de}
\affiliation{University of Osnabr\"{u}ck, Department of Mathematics/Computer Science/Physics, D-49076 Osnabr\"{u}ck, Germany}%
\date{\today}
\begin{abstract}
We suggest a method to compute approximations to temporal correlation functions of few-body observables in chaotic many-body systems in the thermodynamic limit based on the respective Lanczos coefficients. Given the knowledge of these Lanczos coefficients, the method is very cheap. Usually accuracy increases with more Lanczos coefficients taken into account, however, we numerically find and analytically argue that the convergence is rather quick, if the Lanczos coefficients exhibit a smoothly increasing structure. For pertinent examples we compare with data from dynamical typicality computations for large but finite systems and find good agreement if few Lanczos coefficients  are taken into account. From the method it is evident that in these cases the correlation functions are well described by a low number of damped oscillations.
\end{abstract}
\maketitle
\textit{Introduction.}\ 
Understanding out-of-equilibrium quantum dynamics is one of the most fundamental and challenging questions in the field of statistical physics. It has received significant attention, from both theoretical and experimental perspectives\ \cite{polkovnikov11,eisert15,langen15,DAlessio16}.  
Despite the substantial progress that has been made, many fundamental aspects of the problem remain unresolved, and it continues to be the focus of intensive research in the field.

Correlation functions play an essential role in characterizing out-of-equilibrium dynamics and bridge from theory to actual measurable physical phenomena. Over the past decades, the (re)introduction of the Eigenstate Thermalization Hypothesis (ETH) \cite{deutsch91,srednicki94,rigol08,gogolin16} has provided a framework for studying  these functions, particularly for the so-called autocorrelation functions. By postulating a particular structure
of matrix elements of the observable in the eigenbasis
of the Hamiltonian, the ETH explains the essential thermalization of autocorrelation functions after sufficiently long times. However, characterizing their full time evolution lies beyond the scope of the ETH.

\textcolor{black}{To characterize the evolution of correlation functions, on the analytical side, general predictions remain a major difficulty, except for specific cases such as free fermions and certain nontrivial but analytically accessible cases, including particular quantum circuits} \cite{bertini2019,hu2025,hutsalyuk2025}.
In systems with conserved charge, hydrodynamics has been proven to be a useful tool to predict the long-time behavior of autocorrelation functions\ \cite{doyon25,capizzi25,white-PhysRevB.110.134308}. However, it remains an assumption, and the extent of its applicability has yet to be fully established.

On the numerical side, the question is also involved due to the exponential complexity (in system size) of quantum systems. To tackle this question, various sophisticated approaches have been introduced. These range from perturbation theory \cite{Jung2006, Jung2007, Steinigeweg2016}, over quantum typicality \cite{gemmer09,Steinigeweg2014,Heitmann2020}, time-dependent
density-matrix renormalization group methods
\cite{white-PhysRevB.110.134308, Karrasch2012, Karrasch2014} to quantum Monte-Carlo techniques
\cite{Alvarez2002, Grossjohann2010}; each with characteristic advantages and limitations, as well as distinct scopes in terms of application. However, when applied to large system sizes, most of them become quite involved and computationally expensive.

Another approach, the recursion method \cite{mori65,dupuis67,viswanath94}, significantly regained attention thanks to the recent \textit{Operator Growth Hypothesis} (OGH)\ \cite{parker19}. The latter constitutes a statement on the so-called \textit{Lanczos coefficients} of an observable; quantities that fully characterize its dynamics in the thermodynamic limit. Eventually code bases such as \cite{loizeau25-1} position this method along other well-established numerical techniques. Based on the OGH, the recursion method has been applied in various lines fo research\ \cite{NANDY20251}. Particularly it has been applied successfully in the context of correlation functions amounting to studies of dynamical quantities such as equilibration times\ \cite{bartsch24,wang2024-2} and diffusion constants\ \cite{uskov24,wang24}. The success in these applications triggered the question to address the full time evolution of correlation functions in this framework. Recently, in Ref.\ \cite{teretenkov25}, an approach has been suggested to obtain correlation functions via an expansion in terms of pseudomodes, or Ruelle-Pollicott resonances \cite{prosen02,mori_24}, \textcolor{black}{by relying on an extrapolation of the Lanczos coefficients and considering a vanishing artificial dissipation.} 
{\color{black} Besides these endeavors based on the Lanczos coefficients, progress has been made in the computation of correlation functions of hydrodynamical observables including procedures similar to truncated Lanczos expansions, based on tensor networks \cite{PhysRevB.105.075131,PhysRevB.107.L060305,PhysRevLett.132.100402}.}

In our work we put forward an approach to {\color{black}directly} compute correlation functions on the basis of only the \textcolor{black}{practically} numerically accessible Lanczos coefficients\textcolor{black}{, i.e.\ without further explicit assumption on numerical inaccessible ones}. \textcolor{black}{The approach} is both remarkably accurate yet very cheap \footnote{\textcolor{black}{While the cost to compute Lanczos coefficients scales exponentially in the number of coefficients, the first $\sim20-30$ Lanczos coefficients (depending on the model) are readily computable. As will become evident throughout the paper, this number of coefficients typically suffices to apply the scheme.}}. We provide analytical arguments that {\color{black}establish the} applicability of our method in {\color{black}settings with smooth Lanczos coefficients} in accord with the OGH. 
We augment these arguments by numerical investigations of pertinent toy models and specific quantum many-body spin chains, and benchmark using state-of-the-art numerical methods.

\textit{Setting.}\ In this work, we focus on dynamics generated in operator space subject to the recursion method\ \cite{mori65,dupuis67,viswanath94}. Concretely, consider an observable $\vert\mathcal{A})$ regarded as a vector in operator space. The time evolution is implemented by the Liouvillian superoperator $\mathcal{L}=\left[\mathcal{H},-\right]$, where $\mathcal{H}$ denotes the systems' Hamiltonian operator. The Lanczos algorithm orthonormalizes the set of vectors $\{\mathcal{L}^{j}\vert\mathcal{A})\}_{j=0}^\infty$ in a Gram-Schmidt procedure. Concretely, the algorithm is initialized by a normalized \textit{seed} vector $\vert\mathcal{O}_0)=\vert\mathcal{A})/\sqrt{(\mathcal{A}\vert\mathcal{A})}$, where $(\mathcal{A}_1\vert\mathcal{A}_2)=\text{tr}\{\mathcal{A}_1^\dagger \mathcal{A}_2\}/\text{tr}\{\mathbb{1}\}$ denotes the Hilbert-Schmidt scalar product. For $n>0$ it then iteratively computes $\vert\Tilde{\mathcal{O}}_n)=\mathcal{L}\vert\mathcal{O}_{n-1})-b_{n-1}\vert\mathcal{O}_{n-2})$ and normalizes $\vert\mathcal{O}_n)=\vert\Tilde{\mathcal{O}}_{n})/(\Tilde{\mathcal{O}}_n\vert\Tilde{\mathcal{O}}_n)^{1/2}=:\vert\Tilde{\mathcal{O}}_{n})/b_n$, where by convention $b_0=0$. The quantities $b_n$ are the so-called \textit{Lanczos coefficients} and the \textit{Krylov vectors} $\vert\mathcal{O}_n)$ span the \textit{Krylov space} in which the Liouvillian takes tridiagonal form. In order to address dynamical quantities of a system at hand, we define auxiliary functions
\begin{align}
    \varphi_n(t):=i^{-n}(\mathcal{O}_n\vert\mathcal{O}_0(t)),\label{eq_phi}
\end{align}
which inherit from the Liouvillian the discrete Schr\"odinger equation
\begin{align}
    \partial_t\varphi_n(t)=b_n\varphi_{n-1}(t)-b_{n+1}\varphi_{n+1}(t).\label{eq_discreteschroedinger}
\end{align}

For later reference we note that Eq.\ (\ref{eq_discreteschroedinger}) may be cast into a matrix equation
\begin{align}
    \dot{\Vec{\boldsymbol{\varphi}}}=\textbf{L}\cdot\Vec{\boldsymbol{\varphi}},\quad\text{where}\ \textbf{L}_{nm}=b_{n-1}\delta_{n,m+1}-b_n \delta_{n,m-1}.\label{eq_matrixODE}
\end{align}
Imposing the initial condition $\varphi_n(0)=\delta_{n,0}$, solving Eq.\ (\ref{eq_matrixODE}) amounts to the computation of the infinite-temperature autocorrelation function of the observable of interest, as $\mathcal{C}(t)=(\mathcal{O}_0(t)\vert\mathcal{O}_0)=\text{tr}\{\mathcal{O}_0(t)\mathcal{O}_0\}/\text{tr}\{\mathbb{1}\} =\varphi_0(t)$. Conversely, other entries of $\vec{\boldsymbol{\varphi}}$ represent correlation functions between the initial operator and other operators in Krylov space, cf. Eq.\ (\ref{eq_phi}). \textcolor{black}{Note that, for condensed-matter-type settings, frequently $\mathcal{A}$ and $\mathcal{H}$ are local. In this case the first $fL$ coefficients $b_n$ do not depend on system size $L$, with $f$ being a number of the order of one \cite{viswanath94}}.

In principle, solving the matrix equation\ (\ref{eq_matrixODE}) allows for an exact computation of dynamical quantities such as the autocorrelation function, given an infinite amount of Lanczos coefficients. However, typically only a low, two-digit number of $b_n$ {\color{black}pertaining to the infinite system} is numerically feasible, depending on the structure of system and observable alike. To cope with this limitation, the Lanczos coefficients are sometimes extrapolated in accord with the operator growth hypothesis (OGH)\ \cite{parker19}. The latter states that, for few-body observables in chaotic quantum many-body systems that do not have overlap with any conserved quantity, the Lanczos coefficients asymptotically grow linear. Numerically, this behavior has been affirmed in several relevant settings, see e.g.\ \cite{parker19,heveling22-2,noh21,uskov24,wang24}.
It is worth noticing, however, that any truncation of \textbf{L}, albeit at a high dimension, will give rise to \textit{echoes}, i.e.\ unphysical revivals of the correlation function, see e.g.\ \cite{uskov24}.

\textit{Lanczos-Pascal approach.}\  
Concretely, the Lanczos-Pascal (LP) approach {\color{black}(as a recipe)} proceeds as follows. (\textcolor{black}{ Derivations of the applicability of  the LP approach under the below-mentioned conditions on the $b_n$ may be found in End Matter (EM)\ }\ref{app_LP_derivative},\ref{app_theory}.)
As a first step, we determine an area estimation $\mathcal{A}_r $ for $\varphi_0(t)$, i.e., $\mathcal{A}_r \approx  \int_0^\infty  \varphi_0(t)$d$t$ on the basis of the available Lanczos coefficients, given by
\begin{align}\label{eq_area_gm}
    \mathcal{A}_r=\frac{1}{b_1}\prod_{n=1}^{r-1}\left(\frac{b_n}{b_{n+1}}\right)^{(-1)^n}\sqrt{\left(\frac{b_r}{b_{r+1}}\right)^{(-1)^r}}.
\end{align}
This estimate has been suggested in Ref.\ \cite{Joslin86}. It is expected to yield better results with increasing $r$. Let $R$ denote the position after which Eq.\ (\ref{eq_area_gm}) has converged suitably. 
Note that this involves $R+1$ Lanczos coefficients. Next, we infer $D$ as the smallest integer such that $\left(b_{1}/b_{R+1}\right)^D\le\delta$, with some suitably chosen, small $\delta$ (below we employ $\textcolor{black}{\delta=1/10}$). This second condition requires the Lanczos coefficients to grow. With both $R,D$ and $R+D<n_{\max}$ we define 
\begin{align}
\textcolor{black}{s_n}&=\begin{cases}
0 & 0\le n\le R-1\\
\textcolor{black}{(-1)^{n+R+D}\binom{D}{n-R}} & \textcolor{black}{R\le n\le R+D-1}.\label{eq_pascalnumbers}
\end{cases}
\end{align} 
For reasons that shall be explained later, we refer to $D$ as the "order" of the LP curve.  If both above conditions, involving $R$ and $D$, respectively, are met within the accessible $b_n$, we expect the LP approach to yield rather accurate results. This conforms to all examples presented below. 
Note that the nonzero entries in Eq.\ (\ref{eq_pascalnumbers}) are the Pascal numbers with alternating sign, motivating the name of our scheme.

Eventually, we approximate the actual dynamics of $\vec{\boldsymbol{\varphi}}$, as imposed by $\textbf{L}$, by solving the simple matrix equation
 \begin{align}
    \label{eq_approx_ode}\dot{\vec{\boldsymbol{\psi}}}=\textbf{S}\cdot\vec{\boldsymbol{\psi}},\quad\text{where}\ \textcolor{black}{\textbf{S}_{nm}}\textcolor{black}{=\textbf{L}_{nm}+\delta_{n,N-1}b_{N}s_m}
 \end{align}
\textcolor{black}{with} $\vec{\boldsymbol{\psi}}(t)=(\psi_0(t),\psi_{1}(t),\dots,\psi_{N-1}(t))^T$ \textcolor{black}{and} $N:=R+D$ and the initial condition $\psi_n(0)=\delta_{n0}$. Similar approaches to truncations of the Liouvillian have been introduced in \cite{lamann24,loizeau25}. The LP approach  now asserts that $\psi_n(t)\approx\varphi_n(t)$ for $n=0,\dots,N-1$.
Consequently, in analogy to Eq.\ (\ref{eq_phi}), solutions of the simple $N\times N$ matrix equation (\ref{eq_approx_ode}) give rise to approximations of several relevant dynamical quantities ranging from autocorrelations to expectation values in quench settings. Concretely, we have
\begin{align}
    \psi_0(t)&\underset{\text{LP}}{\approx}\left(\mathcal{O}_0(t)\vert\mathcal{O}_0\right)\propto\text{Tr}\{\mathcal{O}_0(t)\mathcal{O}_0\},\label{eq_lp0}\\
    \psi_1(t)&\underset{\text{LP}}{\approx}\left(\mathcal{O}_1(t)\vert\mathcal{O}_0\right)\propto\text{Tr}\{\left[\mathcal{H},\mathcal{O}_0\right](t)\mathcal{O}_0\}\dots\label{eq_lp1}
\end{align}

It is worth noting that by construction, solutions $\psi_k(t)$ are a superposition of at most $N$ damped exponentials. In this sense, LP curves are \textit{simple} for low $N$.

\textit{Numerics.}\ Before turning to physical models, we consider settings in which the connection between the Lanczos coefficients and the analytical form of the autocorrelation function is known. To this end, we study the toy model given by 
$b_n=1+\alpha(n-1)$ for $0\le\alpha\le1$.
In the cases where $\alpha=0$ and $\alpha=1$, the autocorrelation functions are known to be $C(t)=J_1(2t)/t$ and $C(t)=\text{sech}(t)$ respectively\ \cite{Joslin86}. Here $J_n(x)$ denotes the Bessel functions of first kind. The long-time behavior of the former is known to follow a power law, i.e.\ an approximation with \textit{few} exponentials is inherently flawed. We note that in line with our construction, for constant $b_n$, no $D$, as described below Eq.\ (\ref{eq_area_gm}) can be inferred. Here we focus on the two cases $\alpha=0.1,1$ specifically. While for $\alpha=1$ we may compare with the analytical form of the autocorrelation function, for $\alpha=0.1$ we benchmark the LP scheme against the solution of the matrix equation (\ref{eq_matrixODE}) with dimension dim$(\textbf{L})=2000$\ \footnote{As explained in the introduction, solutions of Eq.\ (\ref{eq_matrixODE}) inevitably features \textit{echoes}. Here we therefore only consider times where the correlation function has decayed suitably but well before the echo sets in.}.

\begin{figure}[h]
  \includegraphics[width=0.5\textwidth]{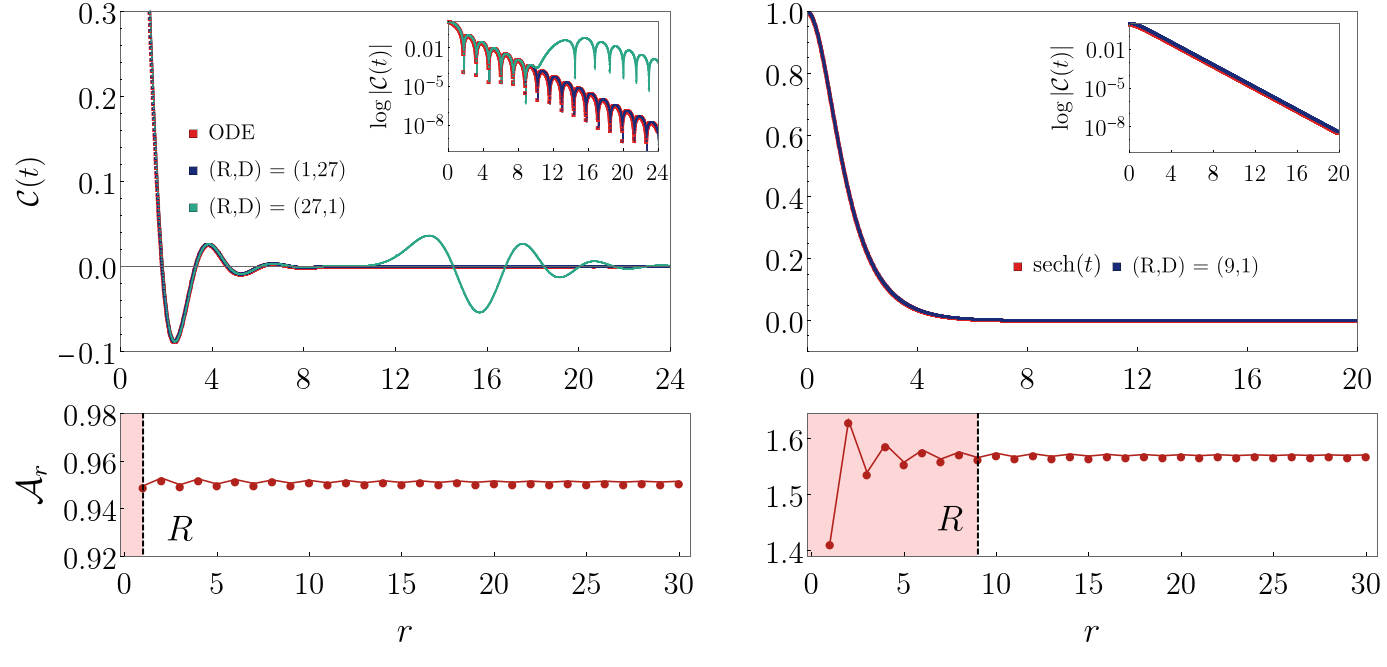}
  \caption{\textit{\textcolor{black}{LP method for a toy model.}} 
  Area estimator $\mathcal{A}_r$ and the autocorrelation function obtained by the LP method and benchmark (inset: log-plot) for $\alpha=0.1$ (left) and $\alpha=1$ (right).
  \label{fig_toymodel}}
\end{figure}

Throughout the paper, we determine $R$ as the position after which $\mathcal{A}_r$ varies less that $1/200\max_j \mathcal{A}_j$ from subsequent areas. In Fig.\ \ref{fig_toymodel} we show the LP scheme at the example of the toy model. Here, the area $\mathcal{A}_{R}$ has converged in the above sense at $R=1$ for $\alpha=0.1$ and at $R=9$ for $\alpha=1$. The quantities $D$ are determined to be \textcolor{black}{27} and 1, respectively. We find that in both cases the LP method gives rise to approximations to the autocorrelation functions with strong agreement to the actual dynamics. For the toy model with $\alpha=0.1$ we also depict the respective LP curve with $(R,D)=(\textcolor{black}{27},1)$ in order for a fair comparison such that in both scenarios the curves consist of the same number of damped oscillations. Unlike for $(R,D)=(1,\textcolor{black}{27})$, the curve eventually fails to capture the true dynamics and deviates visibly indicating the importance of a suitable order in the LP scheme.

\begin{figure*}[t]
  \includegraphics[width=1.\textwidth]{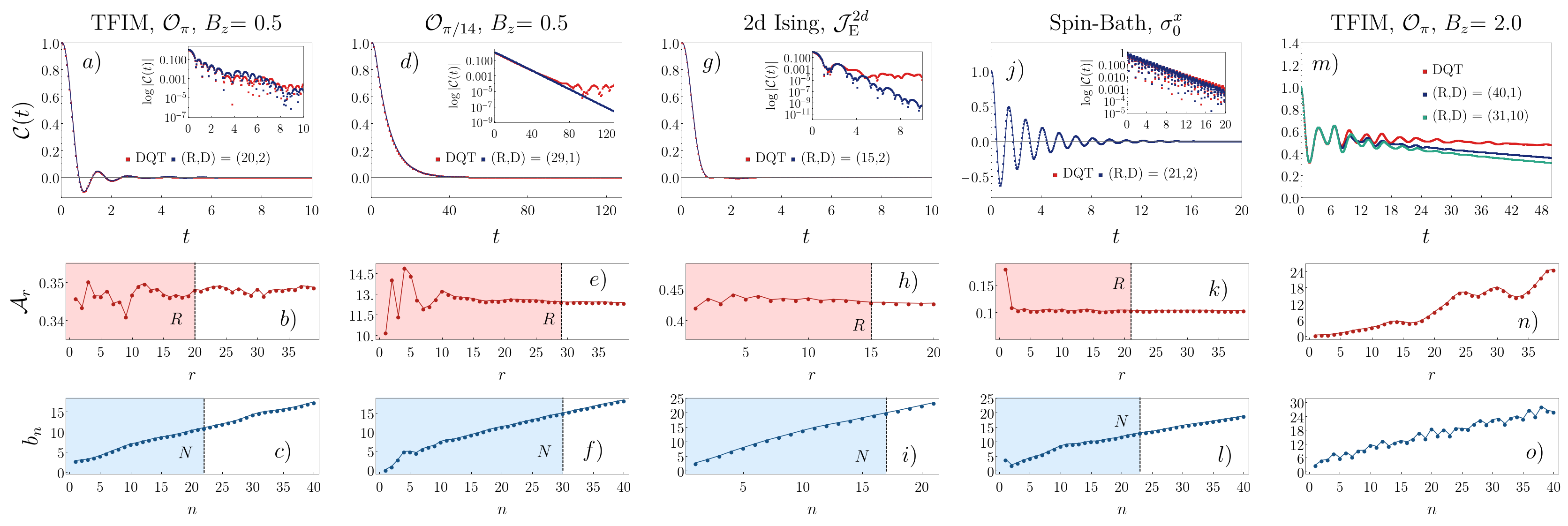}
  \caption{\textit{\textcolor{black}{LP scheme for the autocorrelation function.}} For each system and each observable we depict the respective first $n_{\max}$ Lanczos coefficients $b_n$, the area estimate $\mathcal{A}_r$ along with the point of convergence $R$ and both the autocorrelation function obtained by the LP method and DQT. \textbf{From left to right:} Fast-mode observable in the chaotic Ising chain with $B_z=0.5$, $[(a),(b),(c)]$, slow-mode observable in the chaotic Ising chain with $B_z=0.5$, $[(d),(e),(f)]$, energy current in the two-dimensional Ising chain, $[(g),(h),(i)]$, local $x$ component in the spin-bath model, $[(j),(k),(l)]$, fast-mode observable in the tilted-field Ising model with $B_z=2.0$, $[(m),(n),(o)]$. \label{fig_panel_autocorrelation}}
\end{figure*}
As a first quantum system, we consider as an example
of a generic chaotic quantum system a tilted-field Ising model (TFIM) with the Hamiltonian $\mathcal{H}=\sum_k h_k$, where 
\begin{align}
    h_k=\sigma_k^z\sigma_{k+1}^z+\frac{B_x}{2}\left(\sigma_k^x+\sigma_{k+1}^x\right)+\frac{B_z}{2}\left(\sigma_k^z+\sigma_{k+1}^z\right),
\end{align}
with $B_x=1.05$ and two different field strengths $B_z=0.5,2.0$. In this model, we investigate several different observables, altogether advocating the generality of our findings. In the main part of this paper, we consider density waves of the energy with both short and long wavelengths,
\begin{align}
    \mathcal{O}_q\propto\sum_k\cos (qk) h_k\quad\text{for}\ q=\pi,\pi/14.
\end{align}
Here, we consider up to $n_{\text{max}}=40$ Lanczos coefficients.
{\color{black}See the Supplemental Material \cite{SM} for further examples in the tilted-field Ising model.}
Next, we include into our study a two-dimensional transverse-field Ising model,
\begin{align}
    \mathcal{H}=\sum_{\langle ij\rangle}\sigma_i^x \sigma_j^x+\sum_i \sigma_i^z,
\end{align}
where by the notation $\langle ij\rangle$ the sum is restricted to adjacent sites on the two-dimensional lattice. As an observable of interest we consider the \textcolor{black}{horizontal} energy current
\begin{align}
    \mathcal{J}_E^{2d}=\sum_{\langle ij\rangle\vert i<j} \sigma_i^x\sigma_j^y-\sigma_j^x\sigma_i^y,
\end{align}
\textcolor{black}{with $i<j$ restricting the sum to horizontally neighboring sites.}
This setting has recently been studied in a related context, from where also the first 21 Lanczos coefficients are taken\ \cite{teretenkov25}.
\begin{figure*}[t]
  \includegraphics[width=1.\textwidth]{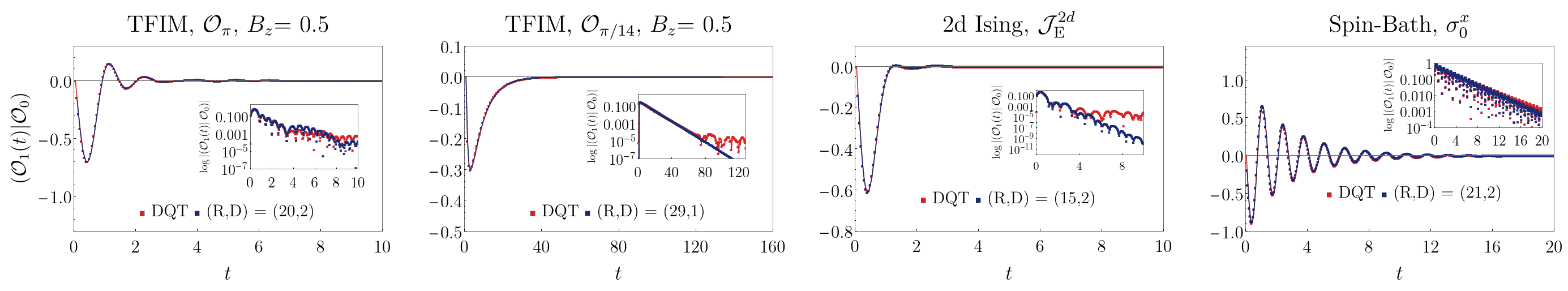}
  \caption{\textit{\textcolor{black}{LP scheme of higher correlators.}} Correlations functions between the first two Krylov states of the respective observable obtained by the LP method and DQT. \textbf{From left to right:} Fast mode and slow mode of the energy density-wave operator in the tilted-field Ising model with $B_z=0.5$, energy current in the two-dimensional Ising model, $x$-component of the local spin in the spin-bath model.
  	\label{fig_higher_correlators}}
\end{figure*}
Lastly, we study an open-system setup, where a single spin is coupled to a large Ising bath. Concretely, we study
\begin{align}
\label{eq_ising_bath}
    \mathcal{H}=\lambda\sigma_0^z\sigma_1^z+\sum_j \sigma_j^z\sigma_{j+1}^z+B_x\sigma^x_j+B_z\sigma_j^z,
\end{align}
with $(\lambda,B_x,B_z)=(2,1,0.5)$. 
Here, the observables of interest is the respective $x$ component of the single spin, i.e.\ $\sigma_0^x$, for which we determine $n_{\max}=40$ Lanczos coefficients.

In order to benchmark the dynamics we obtain from the LP method, we compare with state-of-the-art numerical techniques on the basis of dynamical quantum typicality (DQT)\cite{bartsch09,Heitmann2020}.
For the tilted-field Ising chain, we consider a finite system with length $L=28$, for the transverse Ising model we study a $5\times5$ lattice and for the Ising bath in (\ref{eq_ising_bath}) we have $L=28$. For all systems we employ periodic boundary conditions. The Lanczos coefficients $b_n$ and thus the dynamics generated by the LP approach always concern the thermodynamic limit, i.e.\ $L\rightarrow\infty$.

In Fig.\ \ref{fig_panel_autocorrelation} we depict the autocorrelation functions for the models and observables introduced above. We find that whenever the LP approach is feasible, i.e.\ the area $\mathcal{A}_R$ has converged in the above sense and we may infer $D$, the LP curve is practically indistinguishable from the actual data. It is worth mentioning that this agreement is present, regardless of the feature of the autocorrelation functions, i.e.\ few or many oscillations or mere decay. Deviations set in on a scale that is negligible for most practical purposes. For the tilted-field Ising model with a strong perpendicular field $B_z$, here  $B_z=2.0$, the areas $\mathcal{A}_r$ do not converge within our numerical scope, hence we cannot infer an LP curve subject to the construction introduced above and our $n_{\max}$. Note that, for a magnetic field of this strength, the finite system leaves the chaotic regime, see e.g.\ \cite{wang2024-2}.  To highlight the inapplicablity of the LP method in this scenario, we depict in Fig.\ \ref{fig_panel_autocorrelation} $m)$ two curves, each consisting of the maximal number of damped oscillations allowed by the \textcolor{black}{number of feasible Lanczos coefficients.} Concretely, we show cases where $(R,D)=(40,1)$ and $(31,10)$ respectively\footnote{\textcolor{black}{Note that the parameters $R,D$ for this setting only serve to illustrate the dynamics. They are not inferred with respect to the convergence criteria as described in the main text.}}.

Further, via the LP approach we may address other correlation functions.

For the previous examples with convergent areas, we show in Fig.\ \ref{fig_higher_correlators} the LP approximations $\psi_1(t)$ to \textit{higher} correlators. As with the autocorrelator and $\psi_0(t)$ respectively, the agreement of the LP method is remarkable with deviations only visible on the log-scale.

For all observables that feature smoothly increasing Lanczos coefficients $b_n$, see Figs.\ \ref{fig_toymodel} and \ref{fig_panel_autocorrelation} $[(c),(f),(i),(l)]$, we find the area estimator $\mathcal{A}_r$ to converge within our numerical scope and eventually yielding remarkably accurate approximations via the LP method. Conversely, for cases with \textit{ragged} $b_n$ such as for the fast-mode observable in the TFIM with strong field, see Fig.\ \ref{fig_panel_autocorrelation} $(o)$, no LP curve may be obtained for our $n_{\max}$. 

\textit{Applicability of the Lanczos-Pascal method.}\ 
The applicability of the LP scheme is closely tied to the structure of the Lanczos coefficients $b_n$ themselves, see also EM\ \ref{app_theory}. In order for the LP scheme to apply, the respective areas need to converge. Considering
\begin{align}
    \frac{\mathcal{A}_{r+1}}{\mathcal{A}_r}=
\sqrt{\left(\frac{b_{r}b_{r+2}}{b_{r+1}^2}\right)^{(-1)^r}}\label{eq_smoothness}
\end{align}
we infer that for convergent areas $\mathcal{A}_r$, the Lanczos coefficients need to become sufficiently smooth, i.e.\ $b_{r}b_{r+2}/b_{r+1}^2\approx1$ \footnote{\textcolor{black}{A similar analysis in a related context has already been discussed in Ref.\ \cite{Joslin86}}}. 
Revisiting the Lanczos coefficients for the settings considered in this paper in Fig.\ \ref{fig_panel_autocorrelation} and Fig.\ \ref{fig_higher_correlators}, precisely the cases with smoothly increasing $b_n$ each give rise to accurate descriptions by means of the LP method. \textcolor{black}{Moreover, the \textit{smoothness} of the increase appears to be present in the vast majority of scenarios, while features causing (\textcolor{black}{exceedingly}) high $R$ or $D$ such as staggering or absence or delay of growth of the $b_n$, are mainly found in settings outside the scope of the OGH, i.e.\ for observables overlapping with conserved quantities or systems away from full chaoticity, see}\ \cite{parker19,wang2024-2,wang24,uskov24,teretenkov25,heveling22-2,fuellgraf25,noh21}. 

\textit{Conclusion and Outlook.}\  \textcolor{black}{We introduced the Lanczos-Pascal approach as a scheme based on the recursion method to approximate high temperature autocorrelation functions  in quantum many-body systems in the thermodynamic limit. We present readily checkable criteria based on the respective Lanczos coefficients for the applicability of the LP scheme and find the latter fulfilled for a variety of examples featuring chaotic quantum systems and observables that have no overlap with a conserved quantity. We checked the results of the LP approach against data obtained from dynamical quantum typicality and found good agreement. DQT itself is known to produce results that are in good agreement with state-of-the-art techniques, e.g.,  based on matrix product states etc.\ \cite{Steinigeweg2016}. We do not necessarily expect the LP approach to be more  accurate or have a wider range of applicability than the above state-of-the-art techniques. The impact of the LP-approach rather roots in two features: $i)$ within the regime of its applicability the computational costs for the LP method are much lower than for other methods, $ii)$ without even performing a concrete calculation, the LP approach establishes that, again within the regime of its applicability, the respective correlation functions are well described  by a low number of exponentially damped oscillations. The latter dynamics are ubiquitous in nature, yet to this date encompassing explanations for this ubiquity are scarce.   }

Conversely, in cases where the criteria are not met and hence the LP method becomes inapplicable, this suggests that the underlying correlation function exhibits more intricate behavior, e.g.\ as seen in \cite{PhysRevB.73.035113,mcculloch2025subexponentialdecaylocalcorrelations}. Exploring the possible extension of the LP approach to such cases is an interesting topic for future study.

\textit{Note added.}\ While this manuscript was in preparation, a related study appeared\ \cite{loizeau25}.

\textit{Acknowledgement.}\ We thank Christian Bartsch, Robin Heveling and Mats Lamann for fruitful and valuable discussions. This work has been funded by the Deutsche Forschungsgemeinschaft (DFG), under Grant No. 531128043, as well as under Grant No. 397107022, No. 397067869, and No. 397082825 within the DFG Research Unit FOR 2692, under Grant No. 355031190.

\textit{Data availability.} The data that support the findings of this article are openly
available\ \cite{fullgraf_2025_17733620}.

\bibliography{main}

\begin{figure}[h]
  \includegraphics[width=0.45\textwidth]{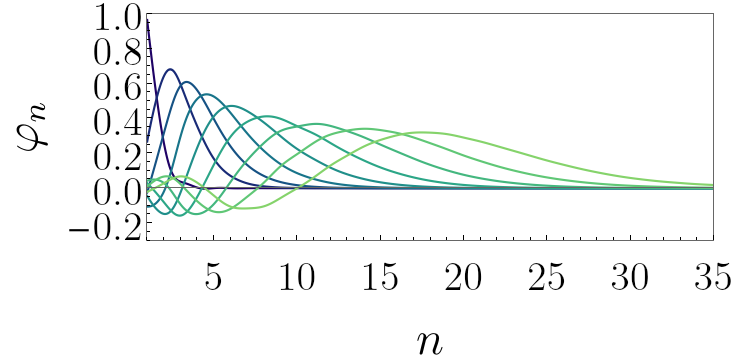}
  \caption{\label{app_fig_wavevector}Wave vector $\vec{\boldsymbol{\varphi}}$ related to the fast-mode of energy-density wave in the tilted-field Ising model with $B_z=0.5$ for various instances of time. Differently to the analysis in the main text, for simplicity we resort to solving the matrix equation (\ref{eq_matrixODE}). \textcolor{black}{The curves are interpolated as the $\varphi_n(t)$ are only defined for $n\in\mathbb{N}$.}}
\end{figure}

\clearpage
\begin{center}
    \textbf{END MATTER}
\end{center}
\textcolor{black}{Below we explain (in two Sections) why the LP scheme yields good results under  conditions as stated in the main text. First we derive its applicability under the assumption of "smooth" $\varphi_n(t)$. To define smoothness we employ Pascal numbers to construct discrete derivatives. Second we establish this smoothness again under the conditions stated in the main text.}
\appendix
{\color{black}
\section{Applicability of the  Lanczos-Pascal scheme under the assumption of smooth $\varphi_n(t)$ \label{app_LP_derivative}}
}
The applicability of the Lanczos-Pascal scheme is closely related to the structure of the wave vector $\vec{\boldsymbol{\varphi}}$, \textcolor{black}{see Fig.\ \ref{app_fig_wavevector} for a sketch of a generic example.}

\textcolor{black}{In order to motivate the effective truncation of the Lanczos method, imposed by the Lanczos-Pascal scheme we \textcolor{black}{start by assuming  that there exists  a set of} 
(so far undetermined) coefficients $s_n$ such that}
{\color{black}\begin{align}
  \textcolor{black}{  \sum_{n=0}^{N}s_n \varphi_n(t)=0 }\label{app_eq_sj}
\end{align}}
\textcolor{black}{holds true \textcolor{black}{ at all times $t$.} In the following we drop the arguments for better clarity. Without loss of generality, we may set $s_N=1$. Rewriting yields}
{\color{black}\begin{align}
    \sum_{n=0}^{N-1}s_n\varphi_n+ \varphi_N=0\Longleftrightarrow\varphi_N=-\sum_{n=0}^{N-1}s_n \varphi_n.\label{app_eq-6}
\end{align}}
\textcolor{black}{This amounts to a guess on $\varphi_N$ based on   all $\varphi_n$ with $n<N-1$ . The latter still      }
\textcolor{black}{fulfill the discrete Schr\"{o}dinger equation (\ref{eq_discreteschroedinger}), whereas for $\varphi^{}_{N-1}$ we have}
{\color{black}\begin{align}
 \partial_t \varphi_{N-1}&=b_{N-1}\varphi_{N-2}-b_{N}\varphi_{N}\\
    &=b_{N-1}\varphi_{N-2}+b_{N}\sum_{n=0}^{N-1}s_n\varphi_n.\label{app_eq_schroedinger}
\end{align}}
\textcolor{black}{Hence the Eqs.}\ (\ref{eq_discreteschroedinger}) \textcolor{black}{and} (\ref{app_eq_schroedinger}) \textcolor{black}{form a linear, autonomous set of differential equations for $\varphi_0,\dots,\varphi_{N-1}$. These equations may be cast into matrix form given in Eq.}\ (\ref{eq_approx_ode}). \textcolor{black}{Thus, if Eq. (\ref{app_eq_sj}) would strictly apply, the LP scheme would produce the exact correlation function.}
\textcolor{black}{As in practice we typically never strictly have Eq.\ (\ref{app_eq_sj}), we employ the notation $\vec{\boldsymbol{\psi}}$ for the wave vector generated by $\textbf{S}$ in the main text.}
\textcolor{black}{Eventually, the LP scheme imposes an explicit form for the coefficients $s_n$, where $n=0,\dots,N-1$. The value for $N$ is set by the parameters $R,D$ as $N=R+D$. With these $s_n$ we revisit the left-hand side of Eq.\ (\ref{app_eq-6}). With the choice of the $s_n$ from Eq.\ (\ref{eq_pascalnumbers}) we effectively implement a numerical \textcolor{black}{"derivative of the wave vector $\vec{\boldsymbol{\varphi}}$ with respect to $n$"} via a (possibly higher order) Newtonian difference quotient. The parameters $R, D$ determine both position and order of this derivative as}
{\color{black}\begin{align}
    \sum_{n=0}^{N-1}s_n\varphi_n+\varphi_N=\sum_{n=R}^{N-1}(-1)^{N+n}\binom{D}{n-R}\varphi_n+\varphi_N,
\end{align}}
\textcolor{black}{implementing a spatial derivative of order $D$ at position $R+D/2$. Consulting a typical example in Fig.\ \ref{app_fig_wavevector} suggests that such an approach may be promising and plausible \textcolor{black}{for sufficiently large $R,D$}.}

\textcolor{black}{For the sake for clarity we will consider the setting addressed in Fig.\ \ref{app_fig_wavevector} where of $R=20$, $D=2$ such that}
{\color{black}
\begin{align}
    \{s_n\}_{n=0}^{N-1}=\{\underbrace{0,\dots,0}_{20},1,-2\},
\end{align}
and eventually 
\begin{align}
    \sum_{n=0}^{N-1}s_n\varphi_n+\varphi_N=0=\varphi_{20}-2\varphi_{21}+\varphi_{22}.
\end{align}
}
\textcolor{black}{Consequently, the construction of the LP scheme exploits the spatial smoothness of the respective wave vector $\vec{\boldsymbol{\varphi}}$, effectively realizing a numerical derivative via a Newtonian difference quotient.}

\textcolor{black}{ 
\section{Conditions on the $b_n$ for the smoothenss of the $\varphi_n(t)$\label{app_theory}}
}
\textcolor{black}{In this section, we examine to what extent smoothness of the wave vector can be expected, \textcolor{black}{i.e. to what extend the validity of Eq. (\ref{app_eq_sj}) with the usage of the respective Pascal numbers may be expected.}  
To this end, we consider the Laplace transformation of $\vec{\boldsymbol{\varphi}}(t)$ which we denote by $\vec{\widetilde{\boldsymbol{\varphi}}}(s)$, where $s=\sigma+i\omega$.} \textcolor{black}{It is straightforward but important to realize that the validity of Eq. (\ref{app_eq_sj})  at all times $t$ is equivalent to the validity of its Laplace transformed counterpart $  \sum_{n=0}^{N}s_n \widetilde{\varphi}_n(s)=0    $  at all $ s$. Hence the LP scheme is expected to work if  all   $\vec{\widetilde{\boldsymbol{\varphi}}}(s)$ are  smooth at sufficiently large $n$ in the sense of the previous section.}
\textcolor{black}{First, we turn to the setting where $s=0$, where we can write the Laplace transformation of Eq.\ (\ref{eq_matrixODE}) as}
\begin{align}
    \begin{pmatrix}
        -\varphi_0(0)\\
        -\varphi_1(0)\\
        \vdots
    \end{pmatrix}=
    \begin{pmatrix}
       -1\\
        0\\
        \vdots
    \end{pmatrix}=\textbf{L}\begin{pmatrix}
        \widetilde{\varphi}_0(0)\\
        \widetilde{\varphi}_1(0)\\
        \vdots
    \end{pmatrix},
    \label{app_eq_laplace_ode}
\end{align}
\textcolor{black}{having used that $\varphi_n(0)=\delta_{n0}$. In the following we abbreviate $\widetilde{\varphi}_n(0)$ by $\widetilde{\varphi}_n$. The form of Eq.\ (\ref{app_eq_laplace_ode}) allows to extract $\widetilde{\varphi}_n$ in terms of the Lanczos coefficients, solving the equation "from top to bottom". Concretely, we may rewrite Eq.\ (\ref{app_eq_laplace_ode}) as}
{\color{black}
\begin{align}
    -\delta_{n0}=b_n\widetilde{\varphi}_{n-1}-b_{n+1}\widetilde{\varphi}_{n+1},
\end{align}
}
\textcolor{black}{where $b_0=0.$ Up to a multiplicative constant $\widetilde{\varphi}_0$ we can solve iteratively to find the eventual entries as}
\begin{align}
   \widetilde{\varphi}_{2n}/\widetilde{\varphi}_0&=1,\frac{b_1}{b_2},\frac{b_1 b_3}{b_2 b_4},\dots\\
     \widetilde{\varphi}_{2n+1}&=\frac{1}{b_1},\frac{b_2}{b_1 b_3},\frac{b_2 b_4}{b_1 b_3 b_5},\dots.\label{app_eq_vectors}
\end{align}

{\color{black}
The quantity $\widetilde{\varphi}_0=\int_0^\infty \text{d}t \  \varphi_0(t)$  constitutes the area under the curve of the autocorrelation function and is in itself a well-studied quantity, see e.g.\ \cite{uskov24,wang2024-2,wang24,Joslin86}. Resorting to the estimator $\mathcal{A}_r$ of that quantity first mentioned in Ref.\ \cite{Joslin86} and utilized in the main text \textcolor{black}{(Eq.\ (\ref{eq_area_gm}))}we may \textcolor{black}{work out a rewriting of }the former as
}
\begin{align}
\mathcal{A}_r&=\widetilde{\varphi}_0\sqrt{\left( \frac{\widetilde{\varphi}_{r-1}\widetilde{\varphi}_{r+1}}{\widetilde{\varphi}_r^2} \right)^{(-1)^r}}
\label{app_eq_area_gm}.
\end{align}
{\color{black}
As now after some $r\ge R$ the quantity $\mathcal{A}_r$ converges, i.e.\ $\mathcal{A}_r\approx\widetilde{\varphi}_0$, we infer that in turn the vector $\vec{\widetilde{\boldsymbol{\varphi}}}$ needs to become smooth, so that thereafter, independently of $r$, the expression in the square root in Eq.\ (\ref{app_eq_area_gm}) needs to become suitably close to 1. Note that while the convergence of the estimator $\mathcal{A}_r$ itself has been set into relation with the growth of Lanczos coefficients \textcolor{black}{already in} Ref.\ \cite{Joslin86}, the argument above directly draws a connection to structural properties of the wave vector. 
\textcolor{black}{We now}
\textcolor{black}{turn to non-zero arguments $s=\sigma+i\omega$ of the Laplace transform $\vec{\widetilde{\boldsymbol{\varphi}}}(s)$. Since the Laplace transform exists for Re($s$)$>0$, the $\widetilde{\varphi}_n(s)$ do not have poles in the right half of the complex plane. Hence we may restrict our analysis to the imaginary axis, $s=i\omega$, by the identity theorem. In this setting, the Laplace transform amounts to the Fourier transform. Concretely, for Eq.\ (\ref{eq_matrixODE}) we may write}
\begin{align}
    -\vec{\boldsymbol{\varphi}}(0)=\left(-i\omega\mathbb{1}-\textbf{L}\right)\vec{\widetilde{\boldsymbol{\varphi}}}(\omega),
\end{align}
or similarly
\begin{align}
    i\omega\widetilde{\varphi}_n(\omega)=b_{n+1}\widetilde{\varphi}_{n+1}(\omega)-b_{n}\widetilde{\varphi}_{n-1}(\omega).\label{app_i_omega}
\end{align}
{\color{black}
We now aim to relate this setting to the $\omega=0$ scenario. Therefore we define}
\begin{align}
      \widetilde{\varphi}_n(\omega)=:\underbrace{f_n(\omega)}_{f_n}\underbrace{\widetilde{\varphi}_n(0)}_{\widetilde{\varphi}_n},\label{app_define}
\end{align}
{\color{black}
such that inserting into Eq.\ (\ref{app_i_omega}) yields
}
\begin{align}
    i\omega f_n\widetilde{\varphi}_n&=(f_{n+1}-f_{n-1})b_{n+1}\widetilde{\varphi}_{n+1}\\
    \Leftrightarrow\left(f_{n+1}-f_{n-1}\right)&=i\omega f_n\left(\frac{\widetilde{\varphi}_{n+1}}{\widetilde{\varphi}_n}b_{n+1}\right)^{-1},\label{app_eq_diff_fn}
\end{align}
{\color{black}
having used that the functions $\widetilde{\varphi}_n(0)$ fulfill 
\begin{align}
    0&=b_{n+1}\widetilde{\varphi}_{n+1}-b_n \widetilde{\varphi}_{n-1}\\\Leftrightarrow\widetilde{\varphi}_{n-1}&=\frac{b_{n+1}}{b_n}\widetilde{\varphi}_{n+1}.
\end{align}
Resorting to the arguments above that suggest a smoothening of the wave vector \textcolor{black}{at $s=0$}, we argue that for some $n\ge R$, we have $\widetilde{\varphi}_{n+1}/\widetilde{\varphi}_n\approx1$. In the continuum limit, i.e.\ $f_n\rightarrow f(n)$ and $b_n\rightarrow b(n)$, Eq.\ (\ref{app_eq_diff_fn}) may hence be written as 
}
\begin{align}
    2 f^\prime (n)\approx i\omega f(n)/b(n)
\end{align}
{\color{black}
which allows for the solution}
\begin{align}
   f(n)=\text{exp}\left(\frac{i}{2}\omega\int_{n_0}^{n}\frac{1}{b(n^\prime)}\text{d}n^\prime\right)\label{app_eq_fn}
\end{align}
{\color{black}
to the function relating the settings where $\omega=0$ and $\omega\neq0$, {\color{black}for $n>n_0$ and $n_0\ge R+D$.} 

Here, the influence of the growth of the Lanczos coefficients becomes visible. Given that the Lanczos coefficients $b_n$ grow slowly or even are constant, already small frequencies $\omega$ will result in highly oscillatory behavior of the $f(n)$, compromising the mapping between the $\omega=0$ and $\neq0$ settings. Conversely, for regularly, \textcolor{black}{e.g. \ linearly} growing Lanczos coefficients, the arguments presented for $\omega=0$ can similarly be made for small $\omega$, including oscillatory behavior into our analysis. These lines of reasoning can most prominently be seen at the example of the toy model, see Fig.\ \ref{fig_toymodel}. There, the slowness of the growth of the Lanczos coefficients ($\alpha=0.1$) evidently necessitates a higher value for $D$, as for $D=1$ the respective LP curve clearly fails to capture the actual dynamics, whereas in the quickly growing scenario ($\alpha=1$) already the order $D=1$ suffices to approximate the dynamics.}
\setcounter{section}{0}
\setcounter{figure}{0}
\setcounter{equation}{0}
\renewcommand{\thefigure}{S\arabic{figure}}
\renewcommand{\theequation}{S\arabic{equation}}
 
\clearpage
\newpage

\section*{SUPPLEMENTAL MATERIAL}
\subsection*{Details on dynamical quantum typicality}

In this section, we provide more detail on the numerical method based on the dynamical quantum typicality (DQT), which we use to calculate the autocorrelation functions shown in the main text.

Let us start from a normalized Haar-random state in the Hilbert space
\begin{equation}
    |\psi\rangle = \sum_j C_j |E_j\rangle,
\end{equation}
where $C_j$ are complex numbers whose real and imaginary parts are drawn independently from a Gaussian distribution. 
According to DQT, the autocorrelation function at infinite temperature $\beta = 0$, $C(t)=\frac{1}{D}\text{Tr}[{\cal O}(t){\cal O}]$ can be written as
\begin{equation}
C(t)=\langle\psi|e^{iHt}{\cal O}e^{-iHt}{\cal O}|\psi\rangle+\varepsilon(t).
\end{equation}
The error scales as
\begin{equation}
    \varepsilon(t)\sim\frac{1}{\sqrt{D}}
\end{equation}
with $D$ being the Hilbert space dimension of the system.
If $D$ is sufficiently large, one has
\begin{equation}\label{eq-Ct-typ}
C(t)\simeq\langle\psi|e^{iHt}{\cal O}e^{-iHt}{\cal O}|\psi\rangle.
\end{equation}
The accuracy of Eq.~\eqref{eq-Ct-typ} can be further improved by taking an average over $N_p$ different realizations of Haar random states $|\psi\rangle$, and
\begin{equation}
    \varepsilon(t)\sim\frac{1}{\sqrt{DN_{p}}} .
\end{equation}
Employing an auxiliary state
\begin{equation}
|\psi_{{\cal O}}\rangle={\cal O}|\psi\rangle,
\end{equation}
Eq. \eqref{eq-Ct-typ} can be written as
\begin{equation}
C(t)=\langle\psi(t)|{\cal O}|\psi_{{\cal O}}(t)\rangle,
\end{equation}
where
\begin{equation}
|\psi(t)\rangle=e^{-iHt}|\psi\rangle,\quad|\psi_{{\cal O}}(t)\rangle=e^{-iHt}|\psi_{{\cal O}}\rangle.
\end{equation}
The time evolution of the two states can be
calculated by
iterating in real time, and we use Chebyshev polynomial algorithm in this paper.

\begin{figure}[h]
  \includegraphics[width=0.5\textwidth]{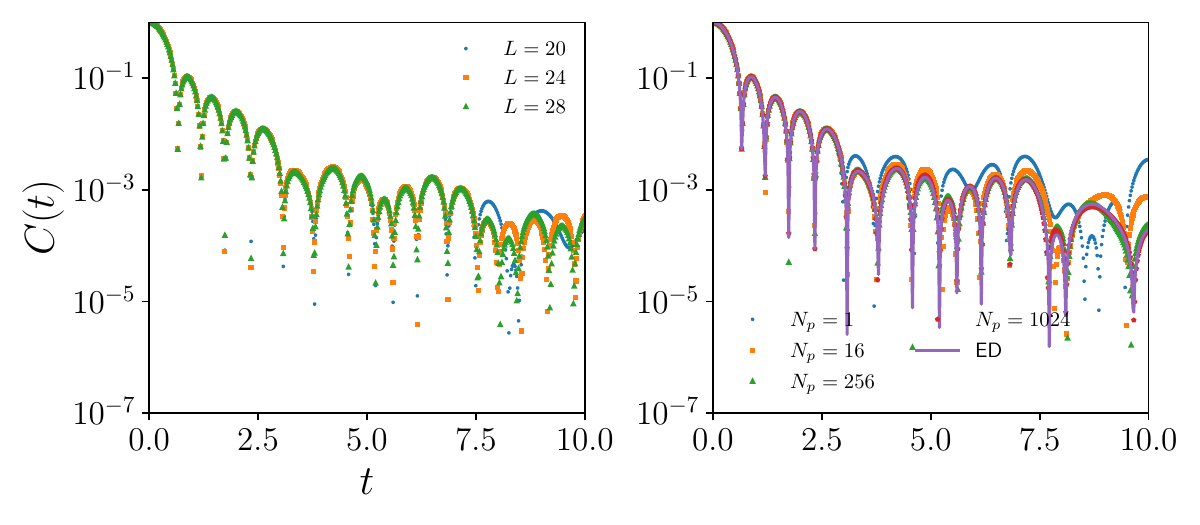}
  \caption{\textit{DQT results for the autocorrelation function of the fast-mode observable $\mathcal{O}_{q=\pi}$ in the tilted field Ising model} (TFIM)\textit{ with $B_z=0.5$.}    \textbf{Left:} Results for different system sizes $L$ averaged over $N_p = 2^{28-L}$ different realizations of Haar-random states. \textbf{Right:} Results for fixed system size $L = 18$ and different $N_p$. The solid line indicates the exact diagonalization (ED) result as comparison. \label{app_fig_typ}}
\end{figure}

\subsection*{Further examples in the tilted-field Ising model\label{app_tfim}}
In the tilted-field Ising model we include further to our study the two-local observable $\mathcal{B}$ and the energy current $\mathcal{J}_E$. Concretely these operators are given by
\begin{align}
    \mathcal{B}&\propto\sum_k J\sigma_k^x \sigma_{k+1}^x-\sigma_k^z,\label{eq_no}\\
    \mathcal{J}_E&=B_x\sum_k\sigma_k^y\left(\sigma_{k+1}^z-\sigma_{k-1}^z\right),\label{eq_jj}
\end{align} with $J=1.05$, for which we compute $n_{\max}=40$ and $n_{\max}=44$ Lanczos coefficients respectively.
In Fig.\ \ref{app_fig_ising} we depict the LP curves for both observables for both $\psi_0(t)$ and $\psi_1(t)$ as well as the area estimator $\mathcal{A}_R$. For the two-local observable $\mathcal{B}$ we find a suitably converged area at $R=18$ and a necessary order $D=3$. For the energy current we have $(R,D)=(35,1)$. Again, the LP scheme gives rise to approximations that are distinguishable from the DQT data to the naked eye. Here, the functions $\psi_0(t)$ and $\psi_1(t)$ both only consist of at most 21 and 36 exponentially damped constituents for the two observables respectively.
\begin{figure}[t]
  \includegraphics[width=0.5\textwidth]{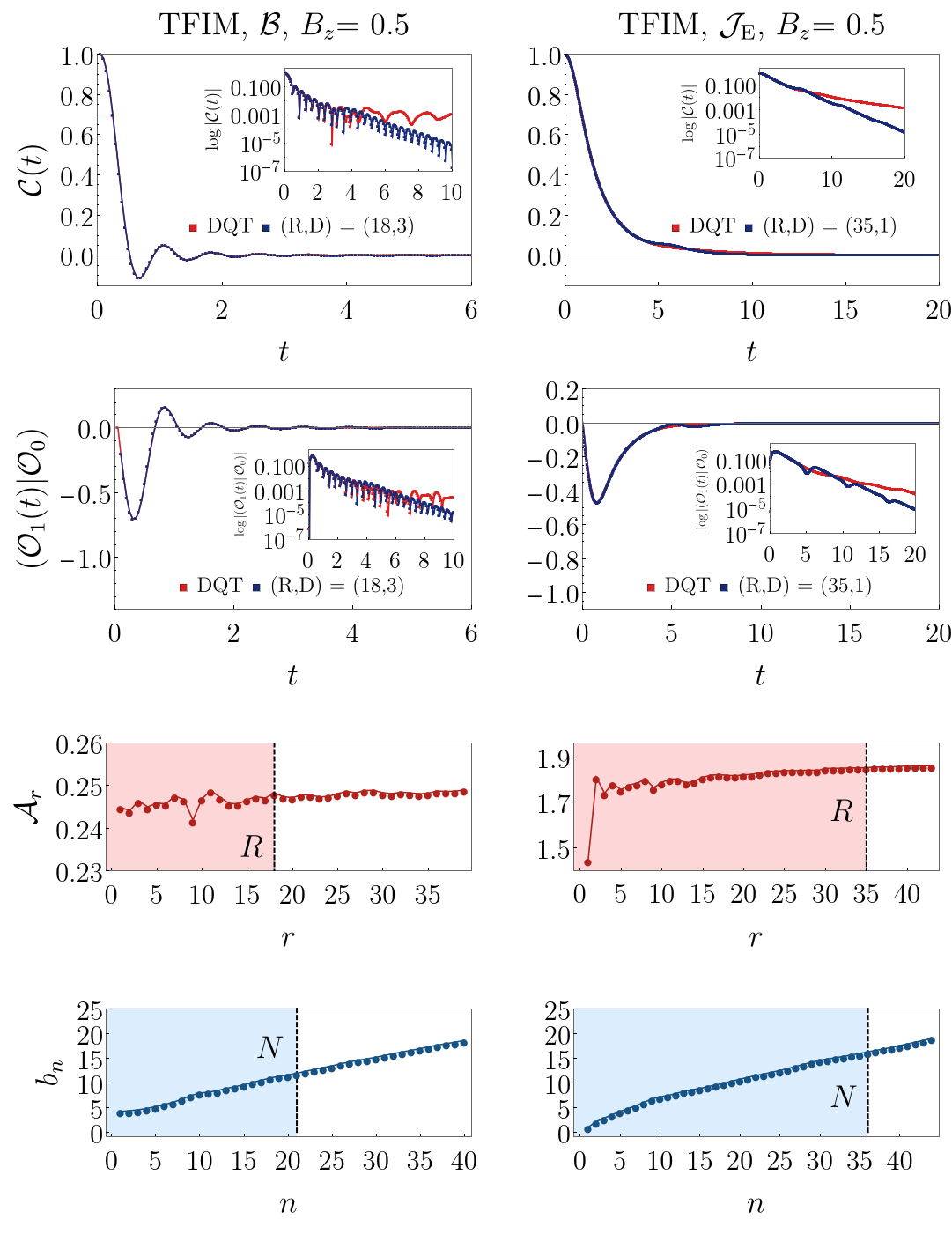}
  \caption{\textit{LP method in the tilted field Ising model} (TFIM). The two-local observable $\mathcal{B}$ (left) and energy current $\mathcal{J}_E$ (right) in the tilted-field Ising model, given by Eq.\ (\ref{eq_no}) and (\ref{eq_jj}) respectively. \textbf{From top to bottom:} Autocorrelation function obtained by the LP method and DQT (inset: log-plot), the correlation between the first two Krylov states of the respective observable, again by means of the LP method and DQT (inset: log-plot), the area estimate $\mathcal{A}_r$ with convergence position $R$ and the respective Lanczos coefficients $b_n$. \label{app_fig_ising}}
\end{figure}

\subsection*{Lanczos-Pascal for transport}
As an application for the Lanczos-Pascal method we compute an approximate diffusion constant for the energy transport in the tilted-field Ising model with $B_z=0.5$ by investigating the low wave number limit of the respective density-wave operator (cf.\ Eq.\ (10)). Concretely, we proceed as follows: we determine up to 40 Lanczos coefficients for wave numbers $q_j=0.01,\dots, 0.5$.
We then apply the LP scheme for each $q_j$ and fit the resulting curve with an ansatz $C_{q_j}(t)=\exp{(-\Gamma_j t)}$, see Fig.\ \ref{fig:fits}. 
We then compare the decay constants $\Gamma_j$ with the respective wave number $q_j$. Performing a further quadratic fit of the former for small $q$ (here: $q=0.01,\dots,0.1$), i.e.\ $\Gamma=\mathcal{D}q^2$, we infer a diffusion constant $\mathcal{D}=1.666$, see Fig.\ \ref{fig:diff_const}. Other analyses in the literature report a diffusion constant in this model $D\approx1.675$, see \cite{parker19,wang24}.
\begin{figure}[h]
    \centering
    \includegraphics[width=0.95\linewidth]{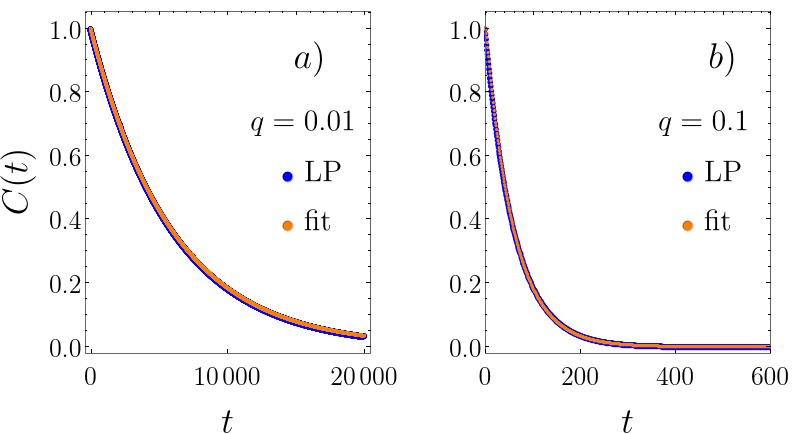} 
    \caption{
    Exponential fits of the LP curves for the energy density-wave operator with wave number $q=0.01$ ($a$) and $q=0.1$ ($b$). The LP parameters are $(R,D)=(25,1)$ and $(23,1)$ respectively. 
    }
    \label{fig:fits}
\end{figure}
\begin{figure}[t]
    \centering
    \includegraphics[width=0.95\linewidth]{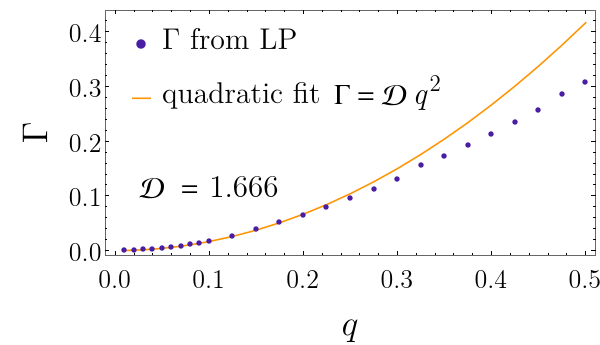}
    \caption{Transport analysis of the LP method. Depicted are the decay constants $\Gamma_j$ from the exponential fits to the LP curves for wave numbers $q=0.01,\dots,0.5$. A quadratic fit of form $\Gamma=\mathcal{D}q^2$ for small $q=0.01,\dots,0.1$ finds a diffusion constant $\mathcal{D}=1.666.$}
    \label{fig:diff_const}
\end{figure}

\vspace{0.1in}

\end{document}